\documentclass[twocolumn,showpacs,preprintnumbers,amsmath,amssymb]{revtex4}

\usepackage{graphicx}
\usepackage{dcolumn}
\usepackage{bm}

\begin{document}


\title{Tweaking the spin-wave dispersion and suppressing the incommensurate phase in LiNiPO$_4$ by iron substitution}

\author{Jiying Li$^{1,2,3}$, Thomas B. S. Jensen$^4$, Niels. H. Andersen$^4$, Jerel L. Zarestky$^1$, R. William McCallum$^5$, Jae-Ho Chung$^{6}$, Jeffrey W. Lynn$^2$, and David Vaknin$^1$\footnote{electronic mail: vaknin@ameslab.gov}}

\affiliation{$^1$Ames Laboratory and Department of Physics and Astronomy, Iowa State University, Ames, Iowa 50011}

\affiliation{$^2$NIST Center for Neutron Research, National Institute of Standards and Technology, Gaithersburg, MD 20899}

\affiliation{$^3$Department of Materials Science and Engineering, University of Maryland, College Park, MD 20742}

 \affiliation{$^4$Materials Research Division, Ris{\o} DTU, Technical University of Denmark, DK-4000 Roskilde, Denmark}

\affiliation{$^5$Ames Laboratory and Department of Materials Science and Engineering, Iowa State University, Ames, Iowa 50011}

\affiliation{$^6$Department of Physics, Korea University, Seoul 136-713 Korea}




\date{\today}

\begin{abstract}
Elastic and inelastic neutron scattering studies of
Li(Ni$_{1-x}$Fe$_{x}$)PO$_4$ single crystals reveal anomalous spin-wave
dispersions along the crystallographic direction parallel to the characteristic
wave vector of the magnetic incommensurate phase. The
anomalous spin-wave dispersion ({\it magnetic soft mode}) indicates the instability
of the Ising-like ground state that eventually evolves into the incommensurate
phase as the temperature is raised. The pure LiNiPO$_4$ system ($x=0$),
undergoes a first-order magnetic phase transition from a long-range
incommensurate phase to an antiferromagnetic ground state at {\it T}$_N$ = 20.8
K.  At 20\% Fe concentrations, although the AFM ground state is to a large extent
preserved as that of the pure system, the phase transition is second-order, and the
incommensurate phase is completely suppressed. Analysis of the dispersion curves using a
Heisenberg spin Hamiltonian that includes inter- and in-plane nearest and
next-nearest neighbor couplings reveals frustration due to strong competing
interactions between nearest- and a next-nearest neighbor site, consistent with
the observed incommensurate structure. The Fe substitution only slightly lowers
the extent of the frustration, sufficient to suppress the IC phase. An energy gap in the
dispersion curves gradually decreases with the increase of Fe content from
$\sim$2 meV for the pure system ($x=0$) to $\sim$0.9 meV for $x=0.2$.
\end{abstract}
\maketitle
\section{Introduction}
Spontaneously occurring incommensurate (IC) structures can be classified into
two general categories. The first group consists of systems for which the IC
phase is the ground state, and the second group encompasses systems for which
the IC phase manifests itself as an intermediate state between a commensurate
ground state and a highly symmetric phase at higher temperatures\cite{Bak1982,
Tsakalakos1984, Mitchell1995}.  Systems with incompatible interactions among
nearest and next nearest neighbors that may lead to {\it geometrical
frustration}, in general belong to the first group settling into an IC ground
state\cite{Bak1982}.  Similarly, nearest-neighbors frustrations brought about
by off-diagonal  Dzyaloshinskii-Moriya type interactions, that compete with the
isotropic interactions, can also give rise to IC ground
states\cite{Dzyaloshinskii1964}.  Magnetic systems consisting of interacting
localized moments, such as MnSi\cite{Moriya1960}, FeGe\cite{Bak1980},
NiBr$_2$\cite{Day1984}, Ba$_2$Cu$_2$Ge$_2$O$_7$\cite{Zheludev1999},
CuB$_2$O$_4$,\cite{Roessli2001}, LiCuVO$_4$\cite{Enderle2005}, and
CdCr$_2$O$_4$\cite{Chung2005} are typical examples of the first group.   On the
other hand, the intermediate IC phases are in general electronically driven by
instabilities due to the incompatibility in the interactions of a collective
mode (phonon) and conduction electrons at the Fermi surface.  These
structurally modulated phases are generally observed in metallic systems as
charge density waves\cite{Moncton1975, Moudden1978}, or martensitic transitions
in alloys\cite{Gooding1989, Noda1990}, and occur at intermediate temperatures
between a disordered state at high temperatures and a highly symmetric ground
state at low temperatures\cite{Mitchell1995}.  Systems belonging to this second
group possess  a few distinct characteristics: 1) they undergo a first-order
commensurate-incommensurate (C-IC) phase transition, 2) they give rise to
strong diffuse scattering above and below the C-IC transition 3) they exhibit
anomalies in their phonon dispersion curves that signal the emergence of the IC
phase.   A typical phonon anomaly appears as a minimum, or a dip, in the
dispersion curve, commonly referred to as a {\it soft-mode}, at a wave-vector
that defines the propagation vector and the shortest wavelength of the IC
modulation.  Due to the first-order nature of the C-IC transition, phonons are
not well defined close to the transition, thus the whole dispersion curve,
including the {\it soft mode}, abruptly disappears near the transition, and in
turn, a {\it frozen phonon} sets in giving rise to a single peak at energy
$\omega \approx 0$.  The {\it frozen phonon}, realized as an elastic or
quasi-elastic peak at and around the wavevector defining the IC structure, is
identified with the IC structure.

An intriguing IC magnetic phase, with features that characterize the second
group, has been found recently in the magnetoelectric crystal LiNiPO$_4$.  The
IC phase occurs over a narrow range of intermediate temperatures between an
antiferromagnetic (AFM) ground state and a high temperature paramagnetic
phase\cite{Vaknin2004}.
Here, it was found that LiNiPO4 undergoes a first order transition from the antiferromagnetic ground state to a long-range IC order at a N{\'e}el temperature,
$T_N$ = 20.8 K ($T_N \equiv T_{C-IC}$).  As the temperature is increased, a
second-order phase transition from long-range incommensurate magnetic order to the paramagnetic state occurs
at $T_{IC}= 21.7$ K\cite{Vaknin2004,Kharchenko2003}.  The incommensurate spin
correlations gradually weaken and the spins are essentially uncorrelated by $T
\approx$ 35 K.   In addition to exhibiting a first order C-IC phase transition,
strong diffuse scattering below and above the transition has also been
reported\cite{Vaknin2004}.  This unusual magnetic intermediate IC phase has
characteristics that classify it with the second group mentioned above,
however, it should be noted that LiNiPO$_4$ is an insulator (with an energy gap
of approximately 1 eV), thus the IC phase cannot be induced by interaction with conduction
electrons.

A recent neutron scattering study\cite{Jensen2009-SW} investigated the spin
dynamics of pure LiNiPO$_4$ to determine the spin Hamiltonian and identify other features that characterize the aforementioned second group of IC systems, particularly looking for a
{\it soft-magnetic mode}, the analog of the {\it soft mode} in structurally IC
systems. An unusual minimum in the spin-wave dispersions in the AFM commensurate ground state was observed at the modulation vector of the IC phase, and was explained as the precursor of the C-IC phase transition that originates from a trade off between competing Heisenberg interactions of nearest and next-nearest neighboring Ni$^{2+}$ ions and an extra {\it lock-in} energy at lower temperatures originating from the strong single ion anisotropies found in the system \cite{Jensen2009-SW}. We have recently reported on the spin dynamics and magnetic
properties of the isostructural LiFePO$_4$, LiCoPO$_4$ and LiMnPO$_4$ systems and found no evidence for an IC
phase and no anomalous spin-wave dispersions \cite{Li2006,Tian2008,Li2009}. In the present study we have substituted Fe for Ni to form LiNi$_{1-x}$Fe$_{x}$PO$_4$ single crystals with up to $x=0.2$
to compare with the magnetic  behavior of the parent material and shed further light on the IC phase in LiNiPO$_4$. Our studies show that, up to a substitution level greater than $\sim 0.15$ iron, the IC
phase is still present, and only at higher Fe concentrations does it disappear
completely\cite{Li-tobepublished}. We report herein the spin
dynamics of LiNi$_{0.8}$Fe$_{0.2}$PO$_4$, that do not show evidence for the IC magnetic
structure, and compare the results with measurements of
pure LiNiPO$_4$.

LiNiPO$_{4}$ is an insulator belonging to the olivine family of lithium
orthophosphates Li$M$PO$_{4}$ ($M$ = Mn, Fe, Co, and Ni) with space group
{\it Pnma}\cite{Abrahams1993}.  All members of this family were found to be  antiferromagnets with the same magnetic structure differing only in the spin-direction\cite{Mays1963,Santoro1966,Santoro1967}, however a recent single crystal study of LiNiPO$_4$ revealed that the magnetic spins are not co-linear in the AFM ground state but are slightly canted within the $ac$-plane \cite{Jensen2009-H}.
Neutron scattering studies demonstrated that  Li$M$PO$_{4}$ ($M$ =Ni,Co,Mn) exhibit
properties between two-dimensional (2D) and three-dimensional (3D) with an
interlayer coupling that is stronger relative to the coupling found in the
cuprates, for instance\cite{Vaknin1999,Vaknin2002,Li2006,Tian2008,Li2009}. These insulators also
exhibit strong linear magnetoelectric (ME) effects, with the observed ME tensor
components, $\alpha_{xz}, \alpha_{zx}$, for LiNiPO$_4$, in agreement with the
antiferromagnetic point groups mm'm, but with some
anomalies\cite{Mercier1968,Mercier1971,Rivera1993}.  In particular, the ME
effect measurements of LiNiPO$_{4}$ as a function of temperature reveal a
first-order {AFM} transition, and an unusual decrease of the ME coefficient at
temperatures below a maximum close to $T_N$\cite{Mercier1968,Vaknin2004}. Recently, a microscopic model combining super-exchange and Dzyaloshinsky-Moriya interactions with elastic displacements of exchange mediating ions has accurately explained the temperature dependence of the ME coefficients in LiNiPO$_4$ \cite{Jensen2009-H}. The model shows that the sharp decrease of $\alpha_{xz}$ and $\alpha_{zx}$ as a function of temperature is intimately connected to the first-order nature of the C-IC phase transition in LiNiPO$_4$. By contrast, the isostructural materials, LiCoPO$_4$, LiFePO$_4$, and LiMnPO$_4$, all exhibit a continuous change of the ME coefficients \cite{Mercier1971} reflecting the second-order nature of their magnetic phase transition from a commensurate AFM state to the paramagnetic state. Magnetic susceptibility studies of
polycrystalline LiNiPO$_4$ showed a significant deviation from the Curie-Weiss
law in a temperature range much higher than $T_N$, and neutron scattering from
the same polycrystalline sample gave rise to diffuse scattering at the nominal
position of the AFM Bragg reflection up to $T \approx 2T_N$\cite{Vaknin1999}.
Recent magnetic susceptibility measurements of single crystal LiNiPO$_4$ showed
two features, one at T$_N$ = 20.8 K  and one at T$_{IC}$ = 21.7 K associated with
an AFM transition and an intermediate IC phase \cite{Kharchenko2003}, in agreement
with the observed neutron diffraction data \cite{Vaknin2004}.

\section{Experimental Details}
LiNiPO$_4$ single crystals were grown by the standard flux growth method (LiCl
was used as the flux) from a stoichiometric mixture of high purity
NiCl$_2$(99.999$\%$) and Li$_3$PO$_4$ (99.999$\%$)\cite{Fomin2002}. To prepare
LiNi$_{0.8}$Fe$_{0.2}$PO$_4$, the Fe substitution was introduced by adding
FeCl$_2$(99.999\%) to the flux at a molar ratio of 1:4 to NiCl$_2$.  The
composition of Li(Ni$_{0.8}$Fe$_{0.2}$)PO$_4$ single crystals were confirmed by
chemical analysis. X-ray diffraction and GSAS refinement show that
Li(Ni$_{0.8}$Fe$_{0.2}$)PO$_4$ has the same crystal structure and symmetry
group as pure LiNiPO$_4$.

The magnetic susceptibility measurements were performed on a superconducting quantum interference device (SQUID) magnetometer. The single crystals used for magnetic property measurements were oriented using Laue back scattering x-ray diffraction. For the different measurements, the single crystals were glued to a plastic straw with the specified axis parallel to the applied magnetic fields with error less than 5$^\circ$.

Elastic and inelastic neutron scattering studies of LiNiPO$_4$ were performed
on the HB1A spectrometer at High Flux Isotope Reactor (HFIR) at Oak Ridge
National Laboratory.  A monochromatic neutron beam of wavelength $\lambda $ =
2.37 \AA\ (14.61 meV, $k_{o}=2\pi /\lambda =2.66$\AA$^{-1}$) was selected by a
double monochromator system, using the (002) Bragg reflection of highly
oriented ( mosaicity 0.3 deg) pyrolytic graphite (HOPG) crystals.   The collimating configuration
40$^\prime $-40$^\prime $-Sample-34$^\prime$-68$^\prime $ was used throughout
the experiments, yielding an average energy resolution of $\approx 1$ meV.  Two
sets of HOPG crystals, located between and after the monochromator crystals,
were used as filters removing the $\lambda/2$ component from the incident beam
to better than one part in 10$^{4}$.  Elastic neutron scattering from
Li(Ni$_{0.8}$Fe$_{0.2}$)PO$_4$ single crystals was measured on the HB1A
spectrometer, and the inelastic neutron scattering was measured on the Spin Polarized Inelastic Neutron
Spectrometer (SPINS) at the NIST Center for Neutron Research (NCNR)
using a fixed final energy of 5 meV.  The collimating configuration
80$^\prime$-Sample-Be filter-80$^\prime$ was used for these measurements
yielding an energy resolution $\approx 0.2$ meV.

\section{Experimental Results}
\subsection{Magnetic Susceptibility}
\begin{figure}
\centering
\includegraphics[width = 0.23\textwidth] {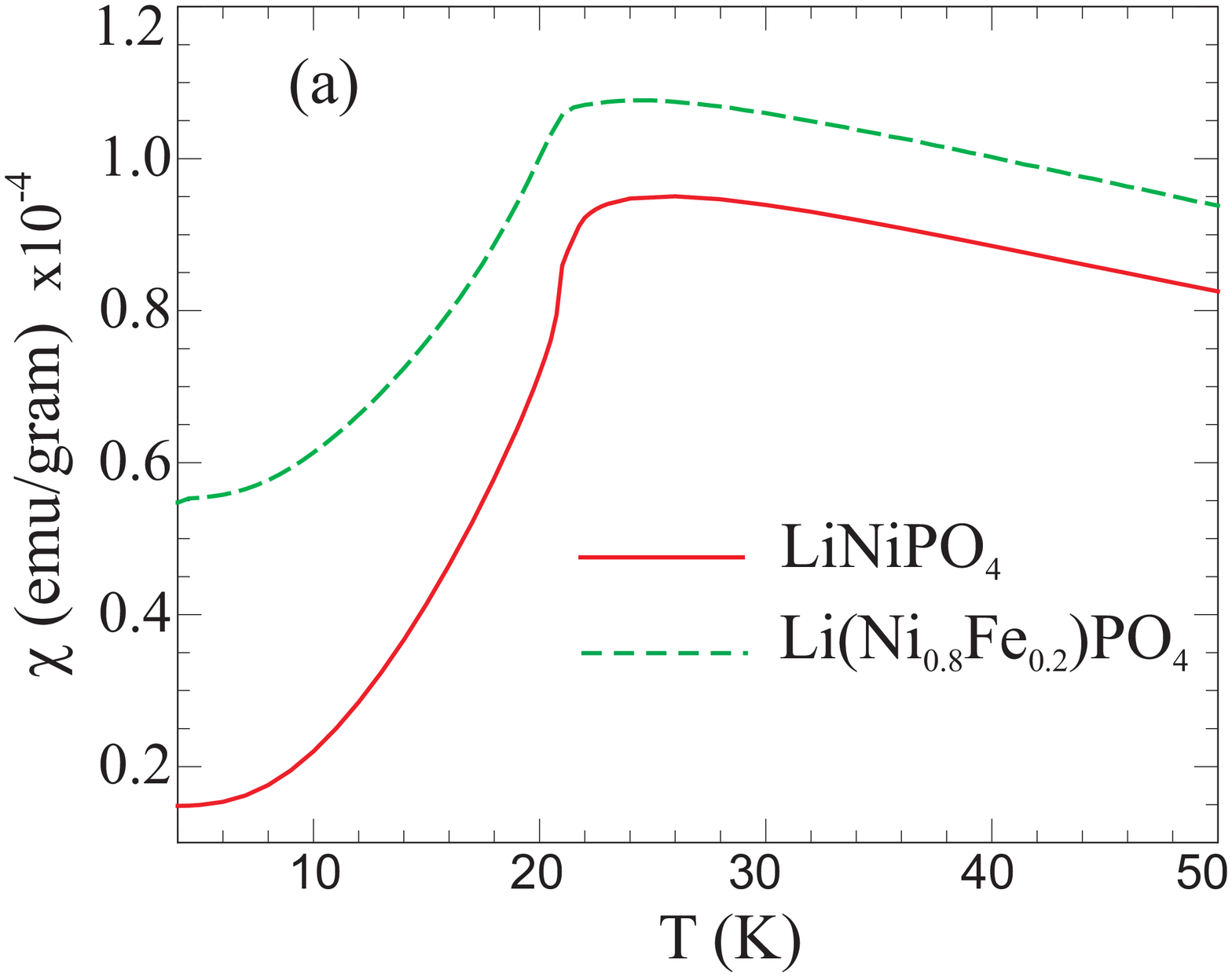}
\includegraphics[width = 0.23\textwidth] {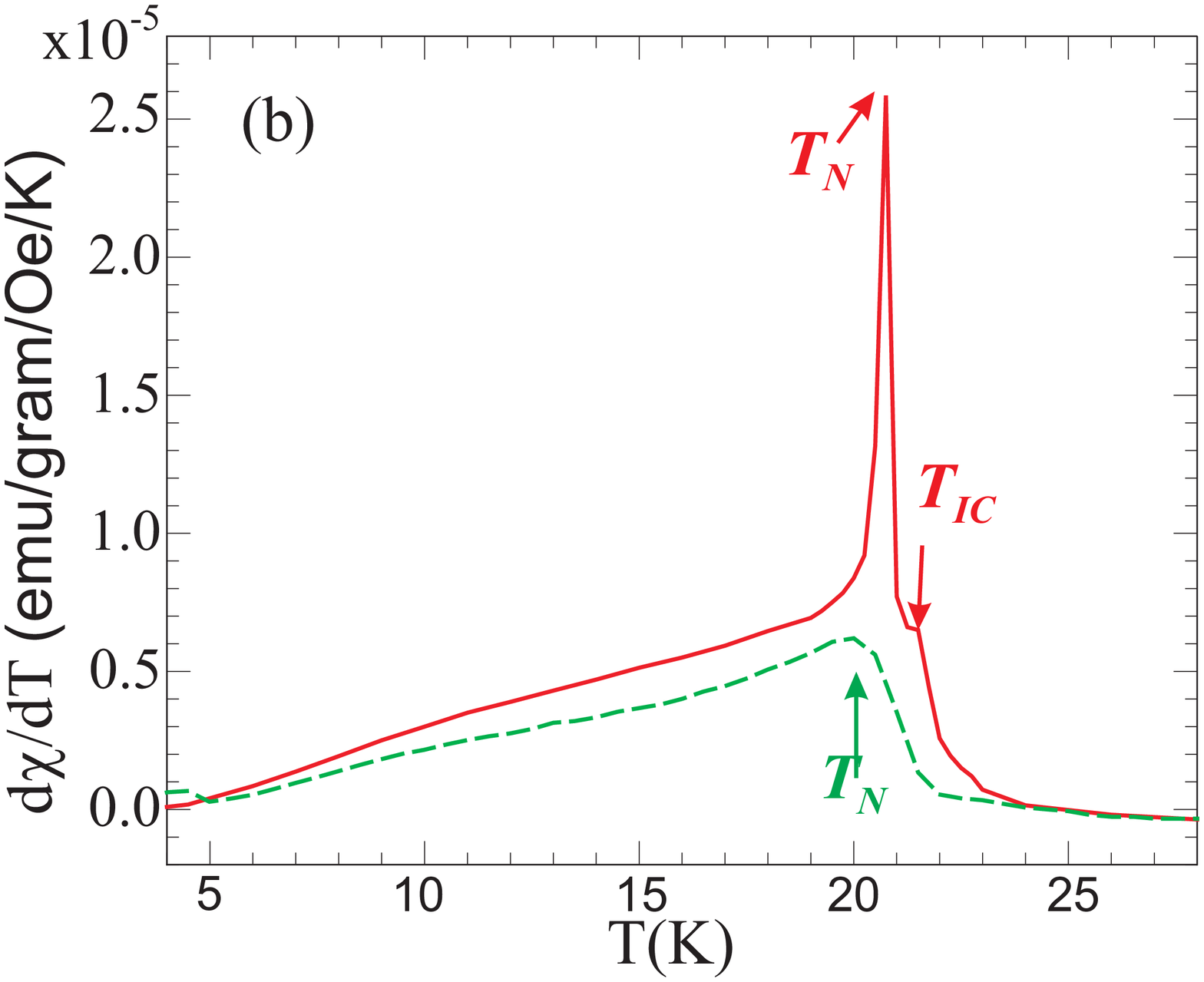}
\caption{(color online) (a) Susceptibility measurements of LiNiPO$_4$ (solid line) and Li(Ni$_{0.8}$Fe$_{0.2}$)NiPO$_4$ (dashed line).  (b) The respective derivatives of the susceptibilities with respect to temperature, showing the two features due to the transitions to IC and AFM in pure LiNiPO$_4$ and a single broad feature (second order transition) in  Li(Ni$_{0.8}$Fe$_{0.2}$)PO$_4$ } \label{chi}
\end{figure}

Magnetic susceptibility measurements of LiNiPO$_4$ and Li(Ni$_{0.8}$Fe$_{0.2}$)PO$_4$ single crystals along the easy c-axis are shown in Fig.\ \ref{chi} (a).  The measured magnetic susceptibilities of the two systems are different in two respects.  First, the absolute value of the susceptibility in Fig.\ \ref{chi} (a) is larger for the iron substituted sample, indicating the presence of uncompensated paramagnetic sites, due to the random distribution of iron spins ($S=2$) with a moment that is different than that of Ni$^{+2}$ ($S=1$).  We note that susceptibility measurements under field- or zero-field cooling indicate subtle spin-glass properties\cite{Li-tobepublished}.   Second, large differences are identified in the derivatives of the susceptibilities with respect to temperature, as shown in Fig.\ \ref{chi} (b).   In agreement with previous measurements of LiNiPO$_4$\cite{Kharchenko2003}, the main AFM-IC transition has the characteristics of a first order transition, and the anomaly associated with the transition from the long-range IC structure to the the paramagnetic state at $T_{IC}=21.7$ K is in good agreement with the neutron diffraction studies\cite{Vaknin2004}.  By contrast the Fe substituted ($x=0.2$) crystal has only one smooth feature characteristic of a second order phase transition with no indication of a secondary transition.  We therefore conclude that whereas the Fe substitution maintains the AFM ground state for all $x \lesssim 0.2$ it does not modify the nature of the transition and does not eliminate the IC long-range order up to a substitution
level of $x \sim 0.2$.
\subsection{Elastic Neutron Scattering}
Neutron diffraction measurements confirm the orthorhombic structure of both the
pure and the Fe substituted samples. For pure LiNiPO$_4$ we find the following
room temperature lattice parameters of $a = 10.030$, $b =5.847 $, and $c = 4.677$ {\AA} and for
Li(Ni$_{0.8}$Fe$_{0.2}$)PO$_4$ we find $a = 10.057$, $b = 5.881$, and $c =
4.672$ {\AA} at 10 K. For the $x =0.2$ sample, we identified a weak
nuclear peak (at room temperature), not identified in the X-ray diffraction of the powder, at the
(010) that may indicate a small structural distortion along the $b-$axis. In
general, elastic neutron scattering and magnetic susceptibility measurements of
the Fe substituted Li(Ni$_{1-x}$Fe$_{x})$PO$_4$ crystals show the
low-temperature ground states of these systems are antiferromagnetic with a
magnetic arrangement similar to that found in pure LiNiPO$_4$.

\begin{figure}[h!]
\centering
\includegraphics[width=0.3\textwidth]{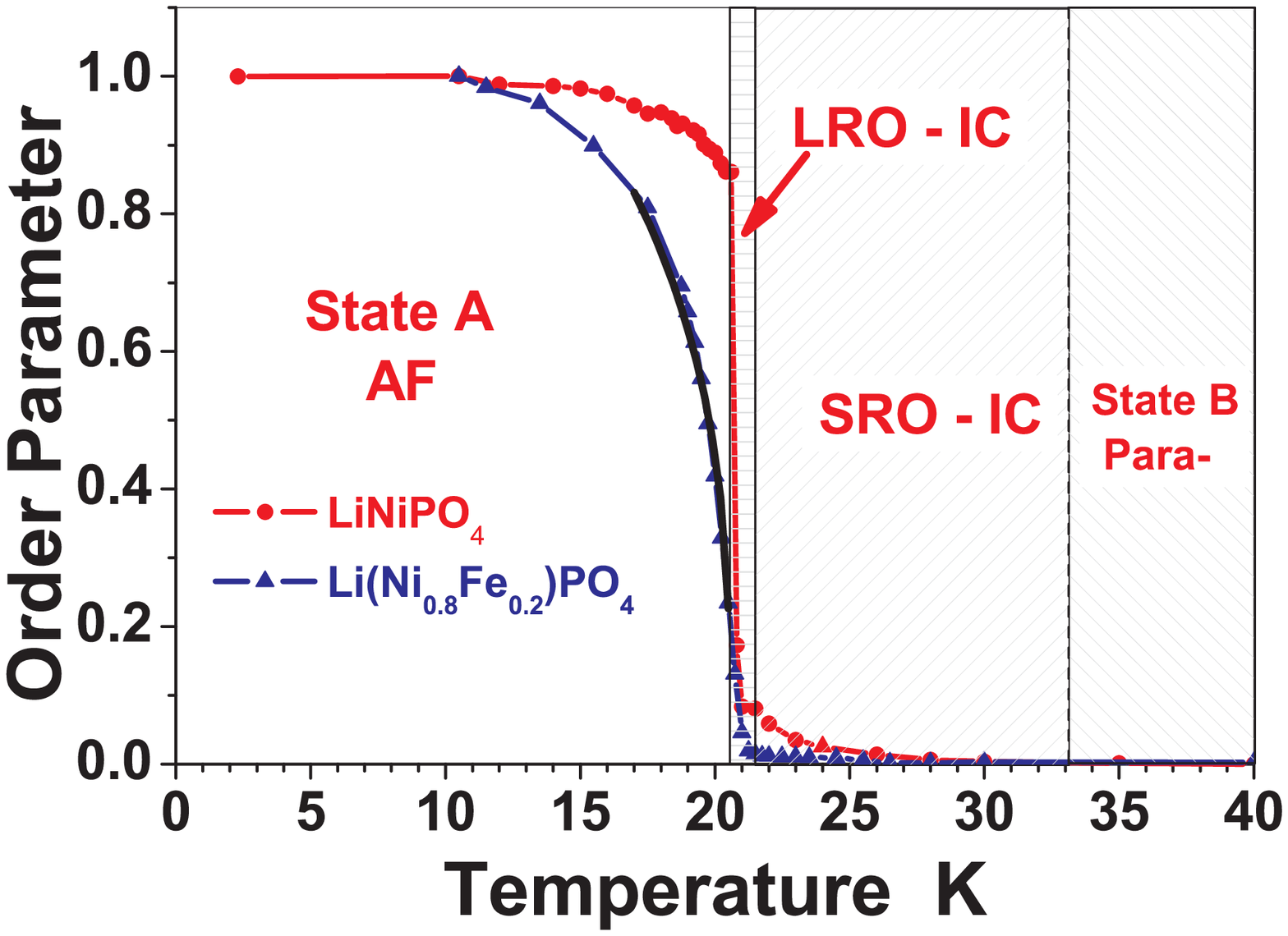}
\caption{(color online) Magnetic order parameter of pure LiNiPO$_4$ and
Li(Ni$_{0.8}$Fe$_{0.2}$)PO$_4$ versus temperature.  LiNiPO$_4$ undergoes a
first order phase-transition from antiferromagnetic ground state to
long-range incommensurate structure at T$_N$ = 20.8 K, and at $T_{IC} \approx
21.7$ K the IC structure transforms to the paramagnetic state. The
incommensurate spin correlations become negligible at about 35 to 40 K. By
comparison, Li(Ni$_{0.8}$Fe$_{0.2}$)PO$_4$ transforms from the collinear ground
state to the paramagnetic state by a second order phase transition at $T_N =
20.6$ K.  The labeled temperature regions refer to phases of pure LiNiPO$_4$.}
\label{order}
\end{figure}

The magnetic spins in LiNiPO4 are primarily directed along the c-axis in the AFM ground state, but are slightly canted with a small component along the {\it a}-axis.  As the Fe concentration is increased, up to at least $x
\approx 0.2$, the N{\'e}el temperature changes slightly but the nature of the
order parameter changes more dramatically.  Figure\ \ref{order} shows the
temperature dependencies of the magnetic order parameters for LiNiPO$_4$ and
Li(Ni$_{0.8}$Fe$_{0.2}$)PO$_4$ as measured on the (010) magnetic peak.   As
previously discussed\cite{Vaknin2004}, LiNiPO$_4$ undergoes a first order
magnetic phase transition from commensurate AFM ground state (labeled A in Fig.
\ref{order}) to a long-range incommensurate structure at $T_{N} = 20.8$ K, and
subsequently to the paramagnetic state at $T_{IC} \approx 21.7$ K by a
second-order phase transition.  The incommensurate spin correlations are
gradually lost by a temperature between 34 K to 40 K.  In contrast, the
transition from the AFM to the paramagnetic phase in
Li(Ni$_{0.8}$Fe$_{0.2}$)PO$_4$ is continuous, i.e., it is a second-order phase
transition to the paramagnetic phase with no clear evidence for any
intermediate magnetic phases.  The temperature dependent order-parameter for
Li(Ni$_{0.8}$Fe$_{0.2}$)PO$_4$, shown in Fig.\ \ref{order}, was fit to a
power-law function (solid line) yielding a transition temperature $T_N = 20.6
\pm 0.2$ K and a critical exponent $\beta = 0.33 \pm 0.03$.
\begin{figure}
\centering
\includegraphics[width = 0.23\textwidth] {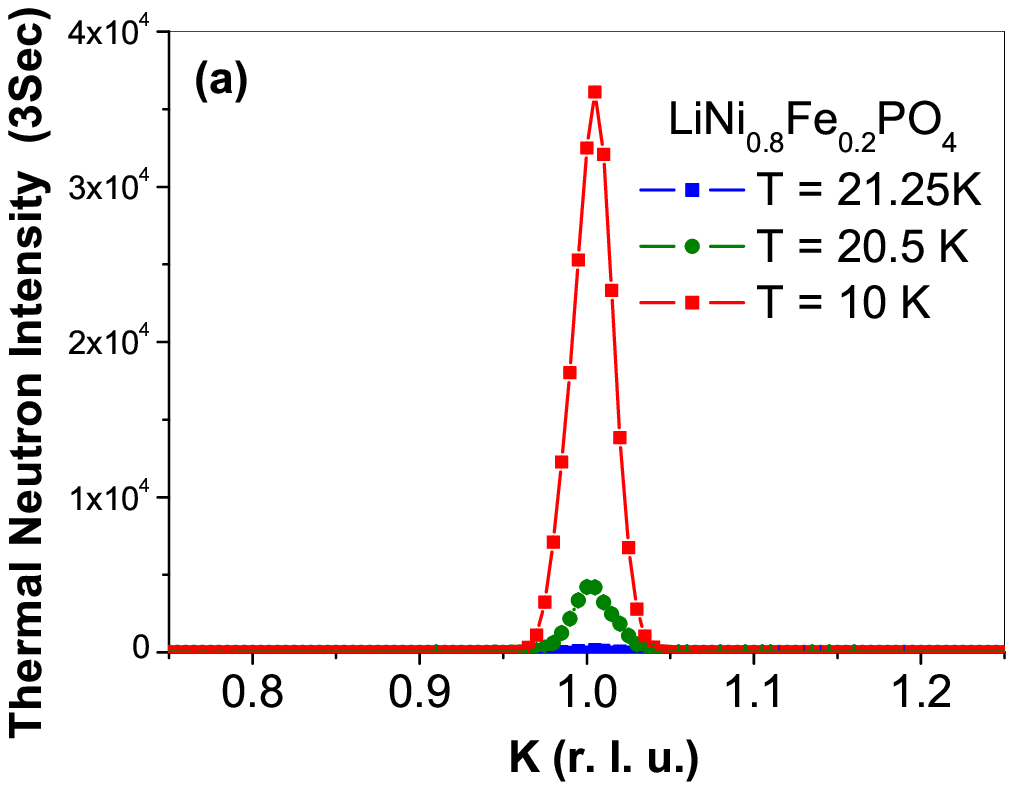}
\includegraphics[width = 0.23\textwidth] {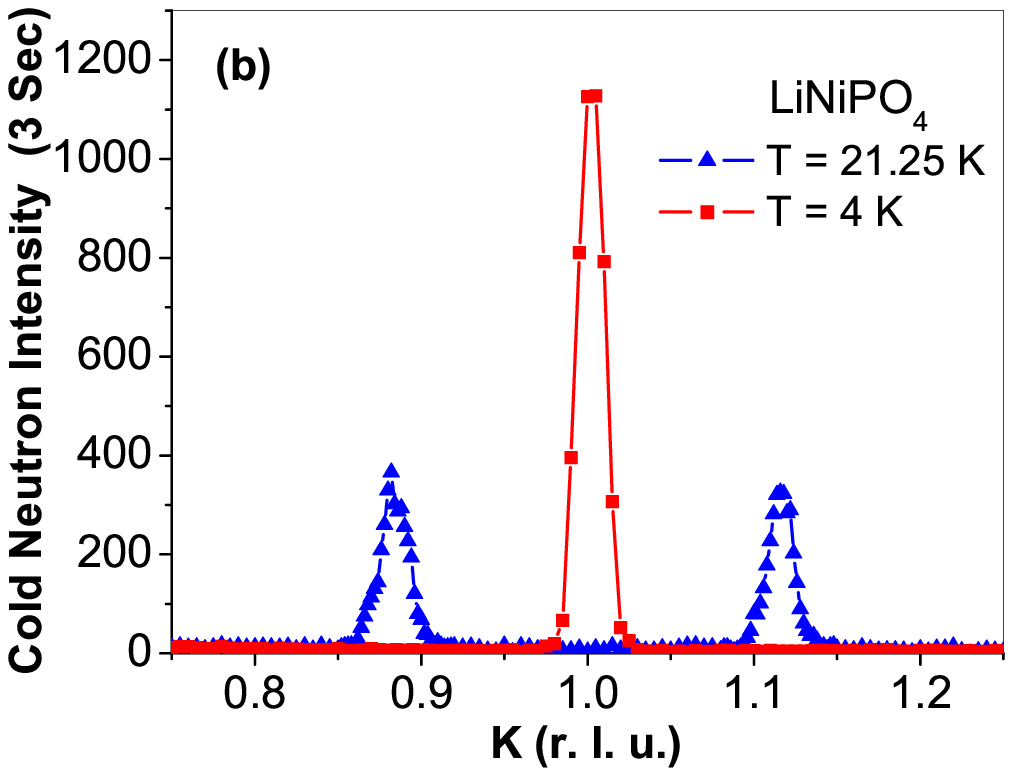}
\caption{(color online) Longitudinal scans along the (0k0) direction for (a)
Li(Ni$_{0.8}$Fe$_{0.2}$)PO$_4$ and (b) LiNiPO$_4$. Above the N{\'e}el
temperature no satellite peaks, due to the long-range incommensurate structure,
were observed for LiNi$_{0.8}$Fe$_{0.2}$PO$_4$ in the temperature range between
18 K to 22 K with 0.25 K temperature steps. The error bars in this paper are
statistical in origin and represent one standard deviation. (r.l.u, stands for reciprocal lattice
units, for example for the (0$q$0) direction $q$ is normalized to $b^* = 2\pi/b$.)} \label{010scans}
\end{figure}

Figure \ref{010scans} shows scans along the (0k0) direction for the pure and
$x=0.2$ samples above the N{\'e}el temperature.  Whereas these scans for
LiNiPO$_4$ above 20.8 K show two satellite peaks due to the IC phase, with
intensities, peak-shapes, and wave-vectors that are strongly temperature
dependent, no similar peaks along the (0k0) direction or along any other
principal direction were detected for Li(Ni$_{0.8}$Fe$_{0.2}$)PO$_4$.

\subsection{Inelastic Neutron Scattering}
\begin{figure}[ht!p]
\includegraphics [width = 0.23\textwidth] {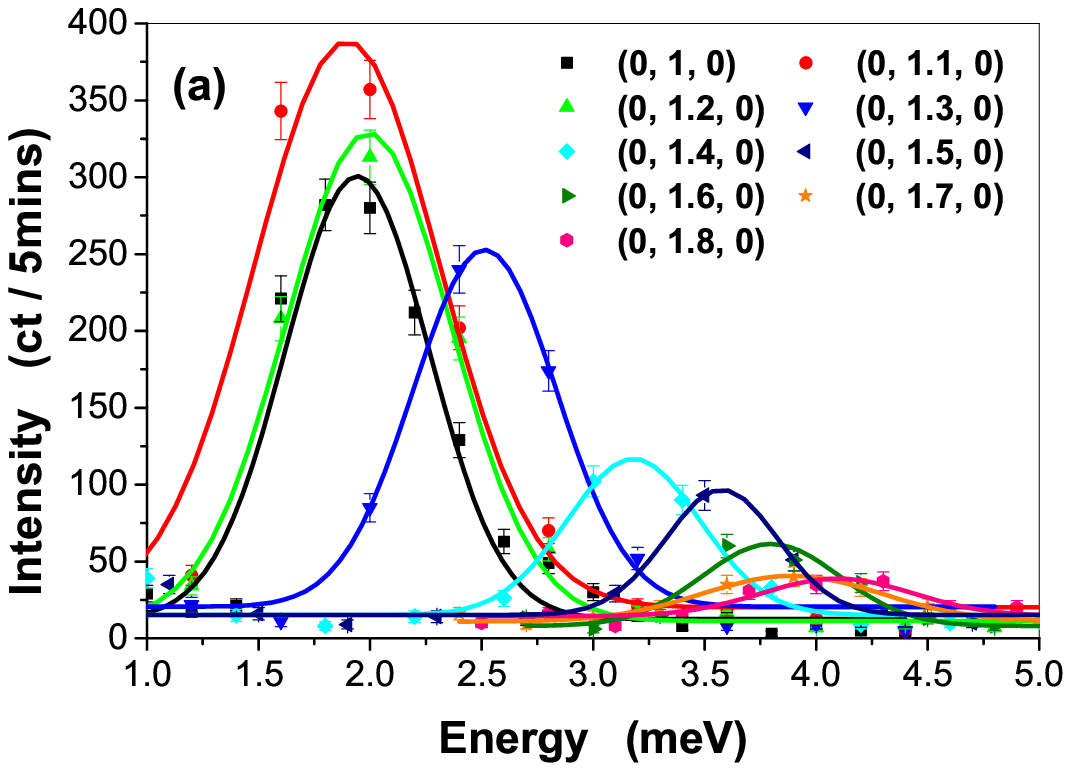}
\includegraphics [width = 0.23\textwidth] {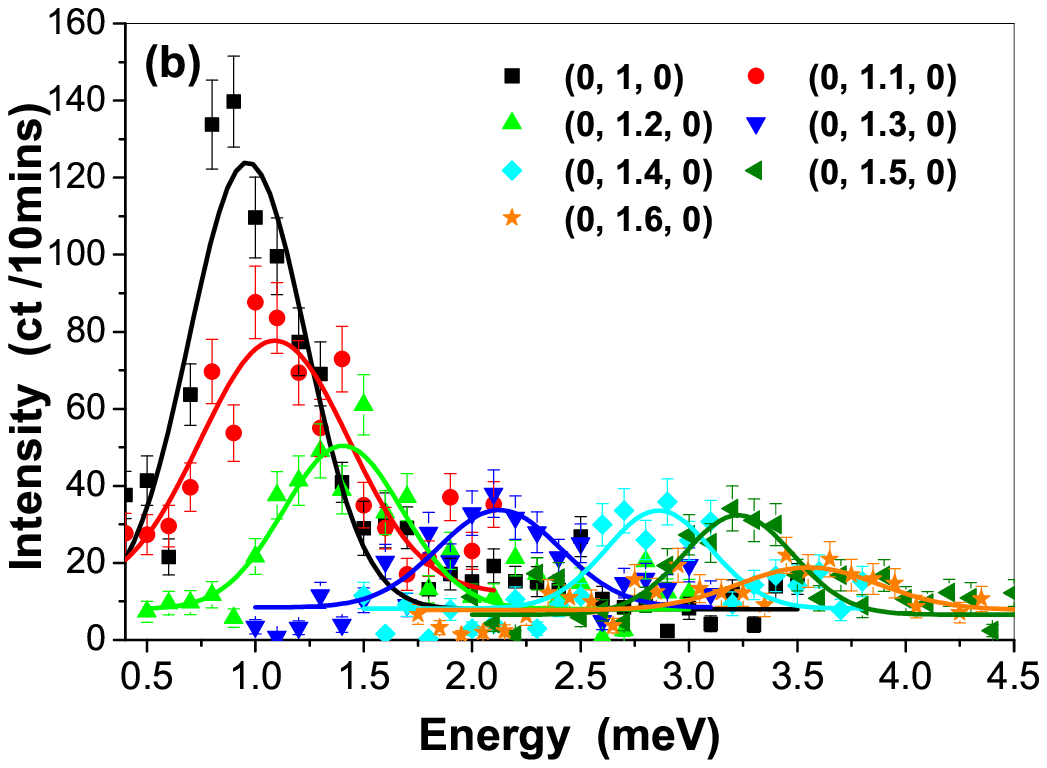}
\caption{(color online) (a) Constant-Q energy scans along the (0$q$0) for a
single crystal LiNiPO$_4$ at 10 K  (b) and for Li(Ni$_{0.8}$Fe$_{0.2}$)PO$_4$
at 4 K. The solid lines are Guassian fits including constant background.  All
modes were measured from the (010) zone center.} \label{EnergyScans}
\end{figure}

\begin{figure}[ht!p]
\includegraphics [width = 0.23\textwidth] {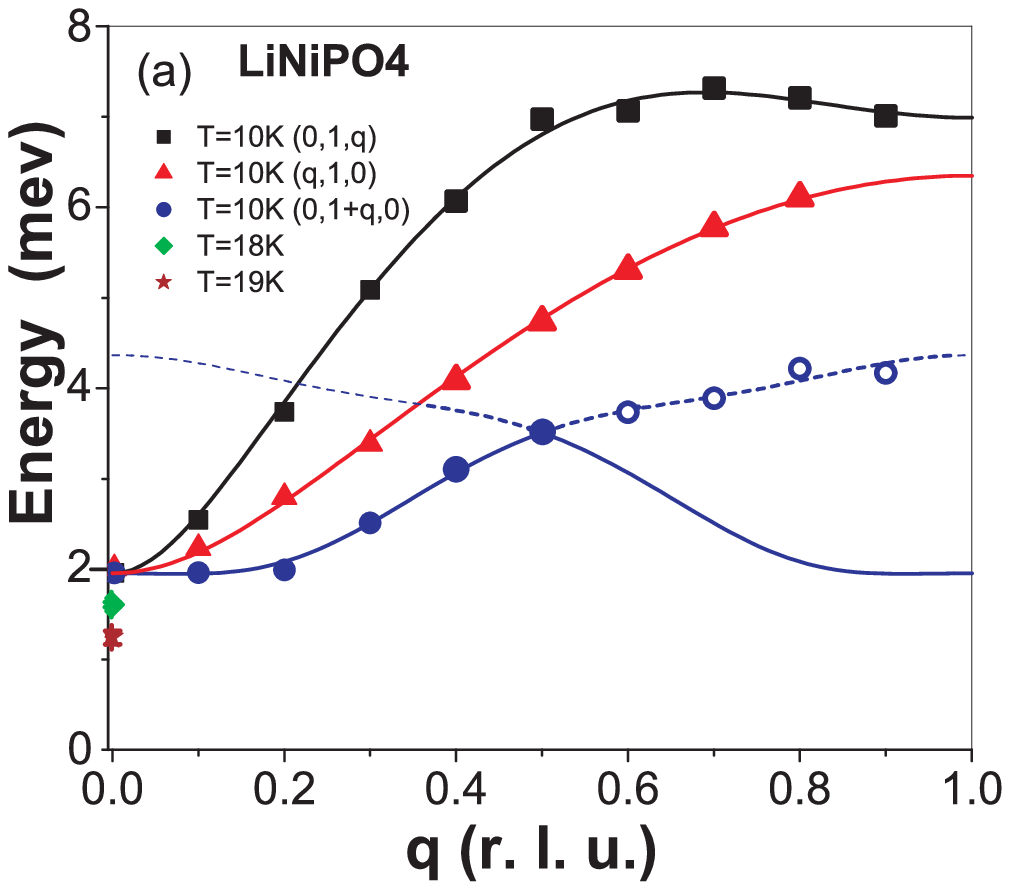}
\includegraphics [width = 0.23\textwidth] {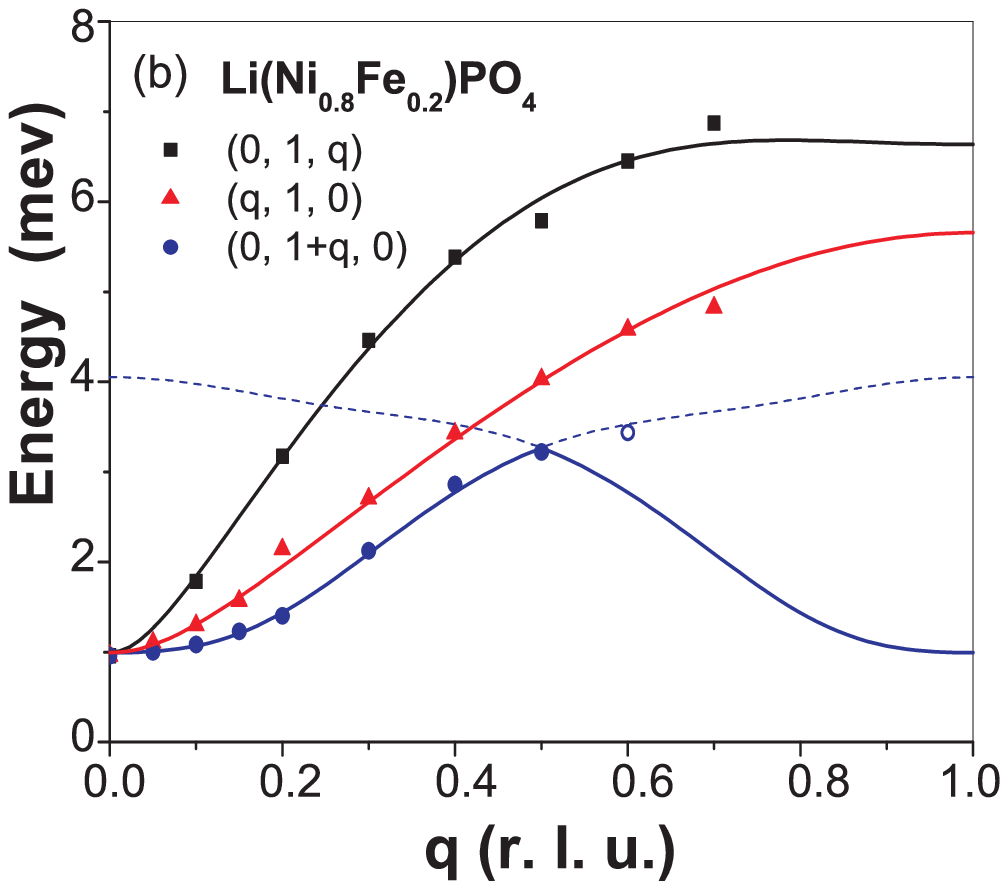}
\caption{(color online) Spin wave dispersion curves along the (00$q$),
(0$q$0) and ($q$00) directions at (a) 10 K for LiNiPO$_4$, (b) at 4 K for
Li(Ni$_{0.8}$Fe$_{0.2}$)PO$_4$.  The solid and dashed lines are fits using the
spin-wave Eq. (\ref{spinwave}). The dotted line starting at the zone center
indicates the spin wave optical branch. }  \label{disper}
\end{figure}

Figure\ \ref{EnergyScans}(a) shows constant-Q energy scans of spin-waves
propagating along (0$q$0) for LiNiPO$_4$ at 10 K measured on the HB1A
spectrometer at HFIR (energy resolution $\approx 1$ meV).  Similar constant-Q
energy scans obtained on the SPINS spectrometer at NCNR (energy resolution 0.2
meV at zero energy transfer) on Li(Ni$_{0.8}$Fe$_{0.2}$)PO$_4$ at 4 K are shown in Fig.\
\ref{EnergyScans}(b).  Each constant-Q scan was fit to a Gaussian profile
(including a constant background) shown as solid lines.  Using this analysis,
the spin-wave dispersion curves along all the three principal directions,
($\xi$00) (0$\xi$0) and (00$\xi$) for both crystals were compiled in Fig.\
\ref{disper}.  The dispersion curves show an energy gap that decreases with
iron substitution.  A gap of $\Delta E \sim$ 1.9 meV is observed for LiNiPO$_4$
compared with $\Delta E \sim$ 0.9 meV for  Li(Ni$_{0.8}$Fe$_{0.2}$)PO$_4$.  The
dispersion curves along the propagation vector (0$q$0) of the AFM structure are
softer (lower in energy) than the curves along the other principal directions.
In particular, it is even softer than inter-layer spin-waves, along the ($q$00)
direction, propagating perpendicular to the $b$-$c$ planes.   This behavior
should be contrasted with the spin-waves of isostructural LiFePO$_4$, where
the dispersion along the (0q0) direction\cite{Li2006} is stiffer than that
along the ($q$00).  Most importantly,  for small $q$'s the curve is almost
flat, with a shallow minimum at $q \approx 0.1$, i.e., a {\it soft magnetic
mode}, whereas for a simple gapless AFM systems the spin-wave dispersion is
expected to be linear at small wave-vectors.  We identify the anomalous
spin-wave dispersion along (0,$q$,0) direction with the {\it soft
magnetic mode}.

The substitution of Fe in Li(Ni$_{1-x}$Fe$_{x}$)PO$_4$ modifies both the energy
gap and the overall dispersion curves.  In particular, the mode along the
(0$q$0) direction is modified and the shallow minimum is not observed, as shown
in Fig.\ \ref{EnergyScans} for Li(Ni$_{0.8}$Fe$_{0.2}$)PO$_4$. These
modifications in the spin-wave dispersion, are not striking considering the
fact that the IC phase is not present in this crystal.  This may suggest that
although the ingredients for the IC phase to occur are still present, namely
competing interactions that lead to frustration, they are not sufficiently
strong or coherent to stabilize an equilibrium IC phase above $T_N$.

\section{Analysis and Discussion}
\subsection{Linear Spin-Wave Theory}
\begin{figure}[t!p]
\includegraphics[width = 0.30\textwidth] {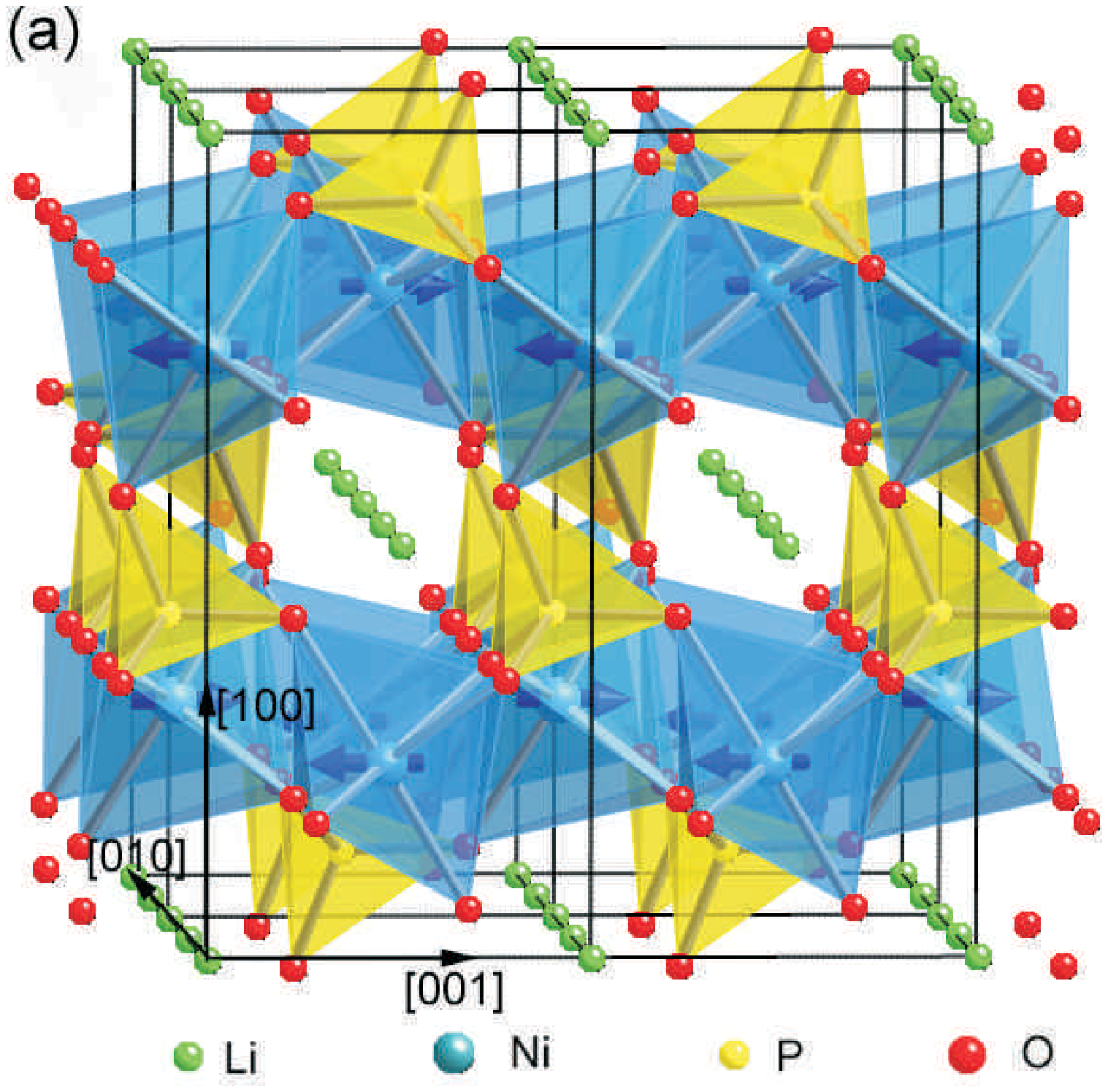}
\includegraphics[width = 0.17\textwidth] {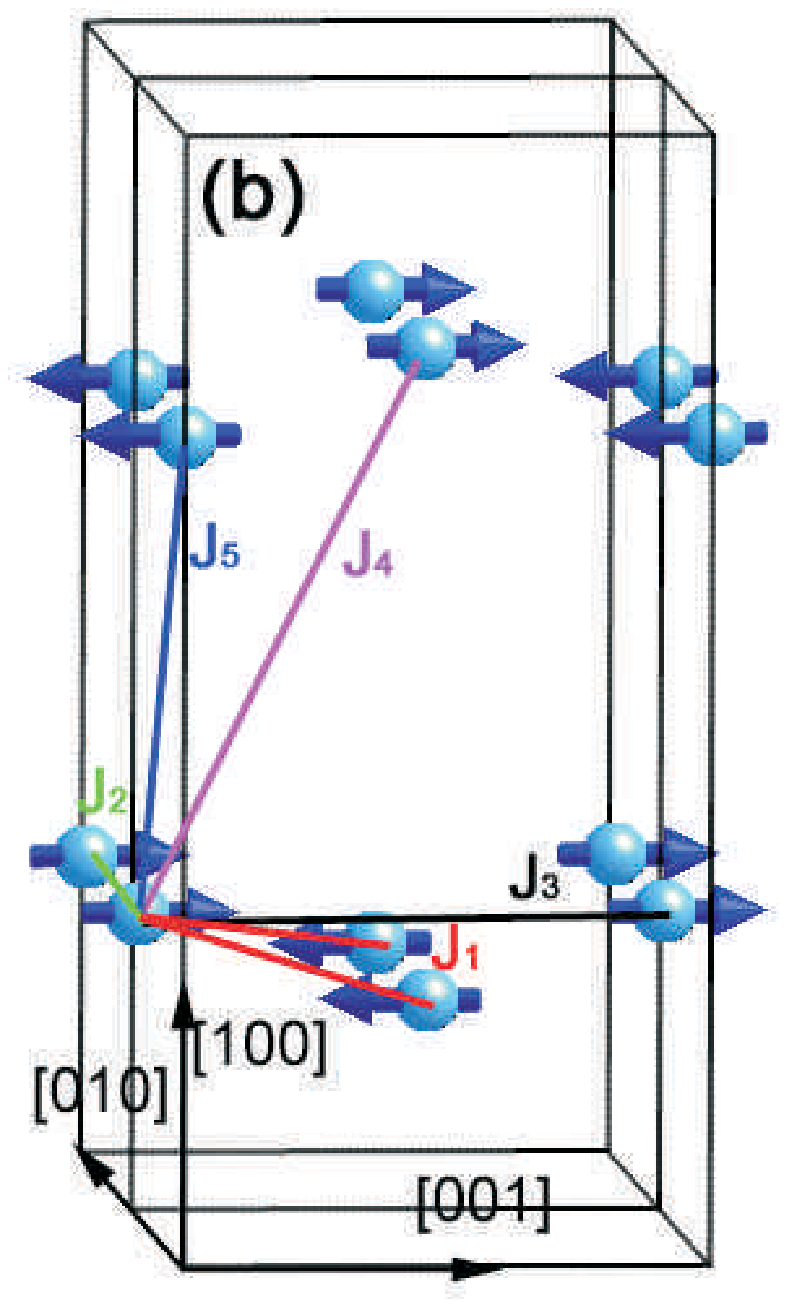}
\caption{(color online) (a) Atomic structure of LiNiPO$_4$. The magnetic
moments of Ni$^{2+}$ are along c-axis in LiNiPO$_4$. (b) Illustration of spin
couplings in LiNiPO$_4$. The same definitions of bonding are used in the spin
wave Hamiltonian} \label{AtomicStr}
\end{figure}

To analyze the measured spin-wave data we follow the model of \cite{Jensen2009-SW} using linear Holstein-Primakoff spin-wave theory \cite{Lindgaard1967,Jensen2009-SW} to calculate the eigenvalues as
a function of wavevectors.  The interaction
parameters determining the eigenvalues are then refined by a non-linear-least-square fit
to the measured dispersion curves. LiNiPO$_4$ adopts the {\it Pnma}
symmetry group, in which Ni$^{2+}$ ($S=1$) ions occupy the centers of slightly
distorted NiO$_6$ octahedra, and P ions are located at the centers of PO$_4$
tetrahedra.  The NiO$_6$ octahedra are corner shared and cross-linked with the
PO$_4$ tetrahedra forming a buckled two-dimensional plane normal to the
$a$-axis.  The atomic structure and definition of spin coupling, used in
this study, are illustrated in Fig.\ \ref{AtomicStr}. The small canting of the Ni$^{2+}$ spins has a negligible influence on the spin wave model \cite{Jensen2009-SW} so for simplicity we have assumed a ground state with spins pointing strictly along the c axis. The inplane
nearest-neighbor (NN) coupling ($J_1$) is mediated by an oxygen through a
Ni$^{2+}$--O--Ni$^{2+}$ bond. The distances between the inplane NN are 3.806
{\AA}.  There are two inplane next-nearest-neighbors (NNN), with distances of
5.891{\AA} and 4.705{\AA}, with inplane couplings $J_2$ and $J_3$,
respectively.  These NNN are linked via Ni$^{2+}$--O--P--O--Ni$^{2+}$ bond. For
inter-layer coupling, we consider only the NN interactions $J_4$ and $J_5$ in
adjacent layers (5.397 and 5.495{\AA} apart, respectively).  The exchange
interaction between NN in adjacent layers is through phosphate tetrahedra.  The
spin-coupling via phosphate tetrahedra can be significant and cannot be
ignored, as has been found for Li$_3$Fe$_2$(PO$_4$)$_3$ where all spins are
coupled via phosphate tetrahedra \cite{Zarestky2001}.
In addition to the Heisenberg interactions, the spin Hamiltonian includes
standard single-ion anisotropy terms $D_{\xi}(S^{\xi})^2 (\xi=x,y,z,)$ as
follows,
\begin{equation}
{\cal H} = \sum_{i,j}(J_{\{i,j\}}\textbf{S}_i\cdot\textbf{S}_j)+ \sum_{i,\xi}D_{\xi}(S^{\xi}_i)^2,
\label{Hamiltonian}
\end{equation}
where $D_{x,y,z}$ are the single ion anisotropies along
the $a$, $b$ and $c$ axis respectively.
{Since the excitation spectrum is insensitive towards an overall shift in the ground state energy we define $D_z \equiv 0$ for simplicity. The magnon
dispersion derived from Eq. (\ref{Hamiltonian}) by linear spin-wave theory is
given by Eq.\ (\ref{spinwave}).
\begin{equation}
\hbar\omega = \sqrt{A^2-(B \pm C)^2},
\label{spinwave}
\end{equation}
where,
\begin{eqnarray}
 A\equiv &&4S(J_1+J_5)-2S[J_2(1-\cos(\textbf{q}\cdot {\textbf{r}}_5))+J_3(1- \nonumber\\
 &&\cos(\textbf{q}\cdot {\textbf{r}}_6))+J_4(2-\cos(\textbf{q}\cdot {\textbf{r}}_7)-\cos(\textbf{q}\cdot {\textbf{r}}_8))]\nonumber\\
 &&+(S-1/2)(D_x + D_y),
\end{eqnarray}
\begin{equation}
B \equiv (S-1/2)(D_x - D_y),
\end{equation}
\begin{eqnarray}
C \equiv &&2J_1S[\cos(\textbf{q}\cdot {\textbf{r}}_1)+\cos(\textbf{q}\cdot {\textbf{r}}_2)]\nonumber\\
&&+2J_5S[\cos(\textbf{q}\cdot {\textbf{r}}_3)+ \cos(\textbf{q}\cdot {\textbf{r}}_4)],
\end{eqnarray}
and $\textbf{r}_i$ denotes a vector to a NN and NNN, $\textbf{r}_1 = (0, b/2,
c/2); \textbf{r}_2 = (0, b/2, -c/2); \textbf{r}_3 = (a/2, b/2, 0); \textbf{r}_4
= (a/2, -b/2, 0); \textbf{r}_5 = (0, b, 0); \textbf{r}_6 = (0, 0, c);
\textbf{r}_7 = (a/2, 0, c/2); \textbf{r}_8 = (a/2, 0, -c/2)$. {In our model, the calculated spin waves have two
non-degenerate branches (denoted by the $\pm$ sign in Eq. \ref{spinwave}) as a
result of the different anisotropies along the $x$, $y$ and $z$ direction.

The energy gaps at \textbf{q} = 0 for the two branches are
\begin{equation}
\Delta E = \sqrt{16S(S-1/2)D_x(J_1+J_5)+4(S-1/2)^2D_xD_y}, \label{EngergyGap}
\end{equation}
for (B - C) in equation \ref {spinwave} and
\begin{equation}
\Delta E = \sqrt{16S(S-1/2)D_y(J_1+J_5)+4(S-1/2)^2D_xD_y}, \label{EngergyGap2}
\end{equation}
for (B + C). The equations show that the energy gaps depending on both the single-ion anisotropy terms and the exchange interactions.

$S = 1$  and $S = 1.2$  are used for LiNiPO$_4$ and Li(Ni$_{0.8}$Fe$_{0.2}$)PO$_4$,
respectively. The experimental data for LiNiPO$_4$ and for
Li(Ni$_{0.8}$Fe$_{0.2}$)PO$_4$ were simultaneously fit for the three principal
directions by Eq. \ref{spinwave} using the (B - C) dispersion. The best fits, shown by
solid lines in Fig. \ref{disper} were obtained by using the parameters listed
in Table \ref{Table1}. It is noted that the values given in Table \ref{Table1} for pure LiNiPO$_4$ are consistent with those reported in Ref.\ \onlinecite{Jensen2009-SW}. The dashed lines in Fig. \ref{disper} are the second
mode of the spin wave calculated using  (B + C) in Eq. \ref{spinwave} and
the parameters listed in Table \ref{Table1}.  For the spin wave dispersion
along the (0, $1+q$, 0) direction, several excitations for the second mode were
measured.  It is clearly shown in Table \ref{Table1} that the inplane NN
exchange $J_1$ is much larger than the inter-plane NN exchanges, $J_4$ and
$J_5$, consistent with the the quasi-2D character of the system. The coupling
constants also show that the NNN inplane coupling along the $b$-axis $J_2$
has the same sign as that of $J_1$, implying competing interactions.  In
particular we find that $J_2$, which couples spins along the $b$-axis, is
significantly larger than $J_3$ that couples NNN along the $c$-axis.  This
is the direction along which the IC structure is realized. The single ion
anisotropies, $D_x$ and $D_y$, are both positive indicating that a $c$-axis
magnetic moment is a favorable ground state, as observed experimentally.   The
Fe substitution systematically weakens all effective spin-couplings and the
single ion anisotropies. Two non-degenerate branches of the spin wave dispersion have
been observed at several scattering vectors in LiNiPO$_4$ using the high-flux thermal neutron triple axis IN8
at Institut Laue-Langevin (ILL) and were perfectly fit by the proposed spin-wave model \cite{Jensen2009-SW}.
\begin{table}[!ht]
\caption{Best fit exchange parameters and single ion anisotropies used to fit the dispersion curves in Fig.\ \ref{disper}.}
\label{Table1}
\begin{ruledtabular}
\begin{tabular} {lll}
 &LiNiPO$_4$& LiNi$_{0.8}$Fe$_{0.2}$PO$_4$ \\
\hline
$J_1$& 0.94(08)& 0.88(15)\\
$J_2$&0.59(05)&0.44(04)\\
$J_3$&-0.11(05)&0.087(02)\\
$J_4$&-0.16(02)&-0.22(04)\\
$J_5$&0.26(02)&0.038(004)\\
$D_x$&0.34(06)&0.072(006)\\
$D_y$&1.92(01)&1.47(1)\\
$D_z$&0&0\\
\end{tabular}
\end{ruledtabular}
\end{table}
\section{Summary}
Model calculations of spin systems with competing interactions between NN and
NNN have demonstrated that anomalous spin-waves, i.e.,
{\it soft-magnetic-mode} are possible for such frustrated systems
\cite{Ivanov93, Liu96}.  The spin couplings for
the Fe substituted compound ($x=0.2$) are slightly different than those of the
the pure one with similar frustrations, but they do not lead to the IC phase.
The realization of the IC phase as an intermediate state may be related to the
energy gap compared to thermal energies at $T_N$.  It is interesting to note
that the energy gap observed in the dispersion curves of the pure system is
very close to $k_BT_N$, i.e,   $\Delta E \approx 1.9$ meV = 22 K.  By contrast
the energy gap in the Fe substituted system is much lower than the intrinsic
$T_N$ temperature $\Delta E \approx 0.9$ meV = 10 K.  Thus, although the
ingredients for the IC phase are present in the Fe substituted sample and give
rise to diffuse scattering, they cannot stabilize the IC structure at any
temperature. Another measure for the feasibility of an IC phase is the ratio of
the competing inplane couplings $J_2/J_1$, for example.  The reduction of the
ratio from $J_2/J_1$ $\approx 0.63$ for LiNiPO$_4$ to $J_2/J_1$ $\approx 0.5$
for the Fe substituted system ($x=0.2$) is sufficient to destabilize the IC
phase.  LiFePO$_4$LiFePO4, with $J_2/J_1 \sim 0.4$, exhibits a second-order paramagnetic-AFM
phase transition with no evidence for the IC magnetic structure at any
temperature \cite{Li2006}.}

In summary, inelastic neutron scattering studies of  LiNiPO$_4$ and
Li(Ni$_{0.8}$Fe$_{0.2}$)PO$_4$ show the spin-dynamics of these systems is
anomalous.  Whereas the anomaly in the pure material leads to an IC intermediate
state, the reduced anomaly in the perturbed system with the substitution of Fe
for Ni does not exhibit an IC magnetic structure.  The spin wave dispersion curves
for both systems were analyzed using the eigenvalues obtained from a
Heisenberg-like spin Hamiltonian by linear spin wave theory.  The spin
couplings obtained indicate frustration between inplane NN and NNN, in
particular along the direction that the IC structure is observed.  Although Fe
substitution does alter the ground state, and preserves the frustration to a
lesser degree, it eliminates the IC phase altogether.
\begin{acknowledgments}
The work at Ames Laboratory was supported by
the Department of Energy, Office of Basic Energy Sciences under contract number DE-AC02-07CH11358.
This work was supported (in part) under the auspices of the United States
Department of Energy. The HFIR Center for Neutron Scattering is a national user
facility funded by the United States Department of Energy, Office of Basic
Energy Sciences- Materials Science, under Contract No. DE-AC05-00OR22725 with
UT-Battelle, LLC.  We acknowledge the support of the National Institute of
Standards and Technology, U.S. Department of Commerce, in providing the neutron
research facilities used in this work which is supported by the National
Science Foundation under Agreement No. DMR-0454672. JHC is supported by the KOSEF grant No. R01-2008-000-10787-0 funded by the Korean government (MEST).
\end{acknowledgments}

\end{document}